# Phase Diagram of Chiral Biopolymer Wigner Crystals

Gregory M. Grason and Robijn F. Bruinsma

*Department of Physics and Astronomy,*
*University of California at Los Angeles, Los Angeles, CA 90024, USA*

(Dated: January 31, 2007)

## Abstract

We study the statistical mechanics of counterion Wigner crystals associated with hexagonal bundles of chiral biopolymers. We show that, due to spontaneous chiral symmetry breaking induced by frustration, these Wigner crystals would be chiral *even if the biopolymers themselves were not chiral*. Using a duality transformation of the model onto a "spin-charge" Hamiltonian, we show that melting of the Wigner crystal is due to the unbinding of screw dislocations and that the melting temperature has a singular dependence on the intrinsic chirality of the biopolymers. Finally, we report that, if electrostatic interactions are strongly screened, the counterions can condense in the form of an intermediate *achiral Wigner solid* phase that melts by the unbinding of *fractional topological charges*.

# I) Introduction

Biopolymers, such as DNA and F-Actin, carry a large negative charge. The resulting electrostatic self-repulsion allows biopolymers to remain soluble in aqueous solution at high concentrations, yet, addition of even very low concentrations of positive polyvalent "counterions" triggers condensation into dense, hexagonal bundles[1,2]. Biopolymer condensation mediated by counterions not only has important applications in biology - in the context of genome compaction - but it also involves fundamental questions of statistical mechanics. The classical mean-field Boltzmann-Poisson theory of aqueous electrostatics only allows *repulsion* between like-charge macro-ions. Biopolymer condensation is thus a consequence of the local correlations and thermal fluctuations that are not included in mean-field theories[3]. Rouzina and Bloomfield[4] first proposed that correlation between polyvalent counterion condensed on a macro-ion could cause them to adopt some form of spatial order, now referred to as a *Wigner crystal*. By suitably staggering the ordered counter-ion arrays of two macro-ions, a strong, short-ranged, electrostatic attraction would result.

This mechanism has been the subject of many analytical[5,6] and numerical[7,8,9] studies. The electrostatic groundstate of an array of neutralizing counterions condensed on a cylindrical surface with uniform opposite surface charge - a primitive model of a biopolymer - is a right or left-handed helical array (see Fig. 1(a)) with a pitch that depends delicately on the range of the electrostatic repulsion[10]. Actual biopolymers like DNA are helical and the pitch of the counterion array would be expected to lock to that of the biopolymer. Simulation studies find that the counterion arrays of two approaching charged cylinders are deformed and that the counterions concentrate along the line of closest contact[7] where they form an array of links ("salt-bridges") connecting the two cylinders (see Fig.1(b)). Thermal fluctuations do not permit long-range positional order for purely one-dimensional systems such as these, so the counterions should be viewed here as a - highly correlated - Wigner *liquid*. Whether the polyvalent counterions of actual biopolymer bundles should be viewed as Wigner liquids or as Wigner crystals has not yet been established[5].

The subject of this paper is the nature of the *melting transition* of Wigner crystals for the case of an infinite bundle of short biopolymers and, specifically, the role of chirality in the melting transition. Figure 2a shows a cross-section of the geometry that is assumed throughout this paper: a dense, hexagonal bundle of biopolymers - indicated by circles - is interspersed by counterion columns placed on lines of close-contact between the cylinders. This array of counterion columns happens to have the symmetry of a *Kagomé lattice*. Counterions can slide up or down these close contact lines and melting is signaled by loss of *phase coherence* between different columns. The melting transition of this simplified model is, we shall show, an elegant and challenging problem of statistical mechanics with implications for other areas of physics such as phase transitions of Josephson Junction arrays.

Figure 2b shows one feature that makes this transition interesting: *frustration*[11]. A triangular prism composed of three adjacent counterion columns is frustrated because the three counterion columns cannot simultaneously minimize their mutual electrostatic repulsion by a pair-wise stagger. The lowest-energy compromise structure is either a left or a right-handed spiral with the adjacent counterion columns out of phase by $120^0$ (see Fig.2b). If one imposes the same choice of chirality on all of the *triangular* prisms, then all of the *hexagonal* prisms of counterion columns surrounding the biopolymers must have the opposite chirality. The pitch of these counterion helices is here fully determined by the geometry of the lattice plus the condition of charge neutrality. The intrinsic chirality of biopolymers breaks the symmetry between the two possible choices for the chirality of the counterion arrays. Importantly, the "$120^0$" phase relation between adjacent counterion chains, which is favored by electrostatics, in general conflicts with the intrinsic pitch of the biopolymers, which forms a second source of frustration.

Given the discrete nature of the broken chiral symmetry, it might seem at first sight that the development of long-range chiral order among the different counterion prisms and the melting transition are entirely separate problems. The former should simply have the character of a two-dimensional (2D) Ising model [12], which has been used to describe other forms of chiral symmetry breaking. One certainly can define an Ising spin variable $S(\mathbf{R}_t) = \pm 1$ on the centers $\mathbf{R}_t$ of each triangular plaquette of the Kagomé lattice that denotes the choice of chirality for that particular plaquette. The intrinsic chirality of the biopolymers would play the role of a "magnetic field" conjugate to $S(\mathbf{R}_t)$ and the electrostatic interaction between adjacent triangular counterion prisms, which favors the same spin choice for adjacent Ising variables, would play the role of an "exchange interaction". The phase-diagram for the development of chiral order would then be isomorphic with that of the Ising model, with a line of first-order transitions terminating at a critical point. Careful consideration of the lattice geometry reveals however that this description is not exact, as most Ising spin arrangements $\{S(\mathbf{R}_t)\}$ are not permitted in the Wigner crystal state. Specifically, one finds that the sum $\sum_{\mathbf{R}_t} S(\mathbf{R}_t)$ around any hexagonal circuit surrounding a biopolymer has to be an integer multiple of three or else the counterion column displacement along the direction normal to the lattice would not be uniquely defined. To satisfy this constraint, *domain walls* separating regions of opposite chirality would have to follow perfectly straight line along one of the three crystallographic directions of the Kagomé lattice, while single chiral spin-flips would be forbidden entirely. Thermal fluctuations of the chiral variable are thus strongly suppressed, at least as compared to the case of the Ising model. On the other hand, loss of phase coherence between the counterion columns would permit a broader spectrum of spin excitations so the melting of the Wigner crystal and the development of long-range chiral order parameter are intrinsically coupled phenomena.

The paper is organized as follows. In Section II we show how the *duality method* can be used to construct a coupled "spin-charge" Hamiltonian. The charges are here topological in nature and mark vortex-like *screw defects* in the spatial pattern of the sliding phase degree of freedom. In this representation, the melting of the Wigner crystal corresponds to the unbinding of *integer* topological charges. In Section III we use the spin-charge Hamitonian to construct the different groundstates and then to show that chiral spin flips and wandering domain-walls with steps and kinks *are* allowed provided we also allow the introduction of *fractional topological charges* $\{Q(\mathbf{R}_h) = \pm 1/3\}$ defined on the centers $\mathbf{R}_h$ of the hexagonal prisms. In Section IV, we combine the Kosterlitz-Thouless (KT) renormalization group equations for both the fractional and integer charge degrees of freedom with a Fisher droplet description for the spin degree of freedom to construct the phase diagram shown in Fig. 3.

In Fig. 3, the coordinate *f* of the horizontal axis is a measure of the strength of the intrinsic chirality of the biopolymers, as discussed in more detail in Appendix B. In the interval $|f|<1$, the counterions of the Wigner crystal locally adopt, in the low temperature phase, the uniform helical-triangular arrangement shown in Fig.2 though "spin-wave" like phase fluctuations produce power-law decay of the positional correlations of counterion order. A line of first-order transitions at $f = 0$, marking spontaneous chiral symmetry breaking, terminates at a *multicritical point* that also is the terminus of two continuous melting transition lines characterized by screw dislocation unbinding. The singular dependence of the melting transition temperature on the chirality parameter is due to the fact that it quenches the domain fluctuations of the chiral order parameter whose characteristic size limits the unbinding of fractional charges. As the range of the electrostatic interaction is reduced, an *achiral Wigner crystal* for a range of intermediate temperatures along the $f = 0$ line. In the achiral Wigner crystal, the counterions are arranged in flat layers perpendicular to the polymers. The achiral Wigner crystal melts by the unbinding of fractional charges. The melting lines are further marked by maxima at $f = \pm 1/2$ while the points $f = \pm 1$ mark the transition, via an intermediate chiral phase with a narrow stability interval, to an achiral Wigner crystal.

## II) The Charge-Spin Hamiltonian

### A) Phase Hamiltonian

The geometry of the model was already shown in Fig.2: a dense bundle of charged, chiral biopolymers of length *L* is placed on a hexagonal grid with a lattice constant *D* close to the hard-core diameter of the polymers, $\sigma_c$. The fixed charges of the polymers have a mean line-density of $-e\rho_0$ and produce an electrostatic potential that is the sum of the radially symmetric potential $\psi_0$ of a line charge plus a multi-pole contribution $\psi_1$ due to non-uniformity in the spatial distribution of the fixed charges. This second term reflects the helical symmetry of the polymer and has a repeat period *p* equal to the pitch of the macroion. Columns of neutralizing counterions of valency *Z* are confined to the array of contact lines that separates neighboring pairs of polymers. The mean spacing $d = 3Z/\rho_0$ between the counterions along the (axial) $z$ direction is determined by the condition of charge neutrality. For $\psi_1 = 0$, the modulation $\delta\rho_i(z)$ of the line charge density of counterion column $i$ around a mean density $Z/d$ would be a periodic function of $z$ with period $d$:

$$\delta\rho_i(z) = \frac{Z}{2d}\sum_{N=1}^{\infty}\left\{A_N \exp(i[Gz - \phi_i]) + cc.\right\} \qquad (2.1)$$

where $G = 2\pi/d$. $A_N$ is here a Fourier coefficient that depends on the structure of the counterions while $\phi_i$ is a phase variable restricted to the interval [0, 2π]. A uniform slide of counterion column $i$ along the $z$ direction by an amount $\Delta z$ would correspond to a phase shift $G\Delta z$. It is assumed in this paper that the length $L$ of the biopolymers is sufficiently short so thermal fluctuations do not produce loss of coherence of the phase variable by phase fluctuations along the $z$ direction. In Appendix A we use the Debye-Hückel theory of aqueous electrostatics to discuss the conditions under which this assumption holds and we also show that the screened electrostatic interaction energy between a pair of neighboring counterion columns $i$ and $j$ is proportional to $\cos(\phi_i - \phi_j)$ so the optimal phase difference of π corresponds to the staggered arrangement of Fig.2b. In Appendix B we show that the chiral contribution $\psi_1$ to the potential produces a shift in the optimal phase difference away from π. The effective "phase Hamiltonian" for the system of phase variables $\{\phi_i\}$ of the different columns is:

$$H[\phi] = V_1 \sum_{\langle ij \rangle} \cos(\phi_i - \phi_j - A_{ij}) + V_2 \sum_{\langle\langle ij \rangle\rangle} \cos(\phi_i - \phi_j) \qquad (2.2)$$

In the first term, $V_1$ is a positive energy scale whose magnitude is determined by the strength of the electrostatic interaction between adjacent counterion columns $\langle ij \rangle$ and by the interaction of the counterions with the chiral potential $\psi_1$. The magnitude of the "bond" variable $A_{ij}$ is a constant that will be denoted by $|A_{ij}| = 2\pi f/3$, with $f$ a measure of the intrinsic chirality of the biopolymers. In Appendix B we show for a simple specific model that if the chiral potential $\psi_1$ is strong, then the repeat period $d$ of the counterion columns "locks" to the pitch of the chiral polyelectrolyte and $f = 1/2$. If the chiral potential is reduced, then the counterion density modulation becomes incommensurate with the helical pitch, and $f$ is reduced as well, while $f = 0$ if $\psi_1 = 0$. The bond variable $A_{ij}$ is odd under exchange of the site indices $i$ and $j$, while the sign of $A_{ij}$[13] is such that if one imposes the optimal phase difference of $\phi_i - \phi_j = \pi + A_{ij}$ on the six counterion columns surrounding a given biopolymer, then this produces a helical charge arrangement that has the same helicity as the biopolymer (see Fig.4, note that the pitch of the counterion helix and the biopolymer in general differ). Finally the second term, with $V_2 \ll V_1$ describes the weak residual electrostatic interaction between pairs of counterion columns located on *next*-nearest-neighbor sites $\langle\langle ij \rangle\rangle$ of the lattice. In Appendix A we show that the interaction between more distance counterion columns can be neglected.

B) Duality Transformations

The Hamiltonian Eq. (2.2) has been studied extensively in the achiral limit $f = 0$ and $V_2 = 0$ in the context of Josephson Junction arrays. We will borrow the method of *duality transformations* developed by Cherepanov et al. [14] for that case to transform our phase Hamiltonian to its dual form, i.e., to a Hamiltonian defined on the centers $\mathbf{R}_h$ and $\mathbf{R}_t$ of, respectively, the hexagons and triangles of the Kagomé lattice (see Fig.5). The starting point is the identity

$$\exp(\beta\cos\phi) = \sum_{n=-\infty}^{+\infty} \exp(in\phi) I_n(\beta) \tag{2.3}$$

with $I_n(x)$ a modified Bessel function. Using Eq. (2.3) one can carry out the integration over the phase variables in the partition function:

$$Z_0 = \int [d\phi] \exp\left\{\beta_1 \sum_{\langle ij \rangle} \cos(\phi_i - \phi_j - A_{ij})\right\} \tag{2.4}$$

with $\beta_1 = V_1 / k_B T$. Note that only the first term of the phase Hamiltonian is included at this point. For every pair $(\phi_i, \phi_j)$ of adjacent phase variables of a bond of the Kagomé lattice, one introduces an integer index $n_{ij}$ that has to be summed over. The phase variables can now be integrated over. Only integer index distributions $\{n_{ij}\}$ contribute that obey the local constraint $\sum_j n_{ij} = 0$ for every site $i$ of the lattice. Next, one defines circulating, integer-valued "currents" $j(\mathbf{R}_h)$ and $j(\mathbf{R}_t)$ on the hexagonal, respectively, triangular dual lattice sites (see Fig.5). The integer summation $n_{ij}$ index is related by $n_{ij} = j(\mathbf{R}_h) - j(\mathbf{R}_t)$ to the integer currents of the hexagon and triangle bordering the $\langle ij \rangle$ bond. The constraints $\sum_j n_{ij} = 0$ are then automatically satisfied so the constrained summations over the set $\{n_{ij}\}$ in the partition function can be transformed into the - more convenient - unconstrained summations over the integer currents:

$$Z_0 \cong \sum_{\{j(\mathbf{R}_h), j(\mathbf{R}_t)\}} \prod_{\mathbf{R}_h} \exp\left(-\sum_{\mathbf{R}_t}' \left\{\frac{\beta_1^{-1}}{2}\left[j(\mathbf{R}_h) - j(\mathbf{R}_t)\right]^2 + 2\pi i \left[j(\mathbf{R}_h) - j(\mathbf{R}_t)\right]\left(\frac{1}{2} + f\right)\right\}\right)$$
(2.5)

The prime in the argument of the exponential indicates summation over the six triangular plaquettes surrounding the hexagonal plaquette at $\mathbf{R}_h$. In deriving Eq. (2.5), it was assumed that $\beta_1$ was large compared to one so the modified Bessel function could be approximated by $I_n(x) \sim \exp(-n^2/2x)$. Finally, one uses the Poisson Summation formula to transform the summations over the integer currents into Gaussian integrals plus summations over integer-valued indices $M(\mathbf{R}_h) = 0, \pm 1, \pm 2,..$ located on the hexagonal plaquettes and odd-integer valued indices $S(\mathbf{R}_t) = \pm 1, \pm 3,.$ located on the triangular plaquettes (details are provided in Appendix C):

$$Z_0 \cong \sum_{\substack{M(\mathbf{R}_h)=0,\pm 1,\pm 2,..\\ S(\mathbf{R}_t)=\pm 1,\pm 3,..}} e^{-H_d/k_B T}$$
(2.6)

The integer indices are now all defined on the dual lattice. The effective Hamiltonian for the integer indices can be written as a sum $H_d = H_b + H_S$ of separate "charge" and "spin" Hamiltonians. The charge part of the Hamiltonian has the form:

$$H_b = \mu \sum_{\mathbf{R}_h} Q(\mathbf{R}_h)^2 - \frac{K}{8\pi} \sum_{\mathbf{R}_h \neq \mathbf{R}_h'} Q(\mathbf{R}_h) Q(\mathbf{R}_h') \ln\left(|\mathbf{R}_h - \mathbf{R}_h'|/D\right)$$
(2.7)

and the spin Hamiltonian has the form:

$$H_S = h \sum_{\mathbf{R}_t} \left(S(\mathbf{R}_t) + 2f\right)^2$$
(2.8)

The parameters of the Hamiltonian all are proportional to $\beta_1$:

$$\frac{\mu}{k_B T} = \frac{\pi^2 \beta_1}{2}$$
$$\frac{K}{8\pi k_B T} = \frac{\sqrt{3}\pi \beta_1}{2}$$
(2.9)
$$\frac{h}{k_B T} = \frac{\pi^2 \beta_1}{6}$$

The quantities $Q(\mathbf{R}_h)$ in Eq. (2.7) are linear combinations of the two groups of integer indices:

$$Q(\mathbf{R}_h) = M(\mathbf{R}_h) + \frac{1}{6}\sum_{\mathbf{R}_t}' S(\mathbf{R}_t) \qquad (2.10)$$

which acts as a local constraint. The summation in Eq.(2.10) is over the six triangular plaquettes that border the hexagonal plaquette at $\mathbf{R}_h$. Finally, the summation over the integer indices also is subject to a global "charge neutrality constraint":

$$\sum_{\mathbf{R}_h} Q(\mathbf{R}_h) = 0 \qquad (2.11)$$

The charge Hamiltonian has the familiar mathematical form of the two-component Coulomb plasma as introduced by Kosterlitz and Thouless to describe the low temperature properties of the 2D XY model [15]. The chemical potential of a topological charge is given by $\mu Q^2$ while $KQ^2$ measures the strength of the 2D Coulomb interaction, *not* to be confused with the DH interaction of the actual Coulomb interaction between the counterions.

The order-parameter correlation function of the phase degree of freedom:

$$g_{ij} = \left\langle e^{i(\phi_i - \phi_j)} \right\rangle \qquad (2.12)$$

is proportional to the correlation function $\langle \delta\rho_i \, \delta\rho_j \rangle$ for the counterion charge density fluctuations. The Fourier transform of $\langle \delta\rho_i \, \delta\rho_j \rangle$ is proportional to the in-plane structure factor $S(\mathbf{q}_\perp)$ of the system of counterions. It is useful to recall here that in the low-temperature phase of the XY model, topological charges only come in bound pairs. In that regime, the correlation function has a power-law dependence on separation:

$$g_{SW} \propto \frac{1}{|\mathbf{R}_{ij}|^\eta} \qquad (2.13)$$

with a - temperature dependent - exponent $\eta = 4\pi k_B T / K$. The decay of the correlation function is due to "spin-wave" like thermal fluctuations of the phase degree of freedom. From the physical interpretation of the phase variables for the case of Wigner crystals, it follows that spin-wave excitations would correspond to *transverse shear modes* of the Wigner crystal with displacements along the direction of the counterion columns and wavevectors in the plane. The shear modulus of the Wigner crystal for displacements along the z direction would correspond here to the phase stiffness $K_{eff}$. In the high temperature phase of the XY model one encounters free topological charges, which correspond to unbound screw dislocations for the case of the Wigner crystal. The area density $n_f \propto 1/\xi^2$ of the free charges determines the correlation length for exponential decay of the correlation function $\langle \delta\rho_i \delta\rho_j \rangle$. The shear stiffness and $K_{eff}$ vanish in this regime, which thus can be identified with a correlated Wigner liquid.

However, as discussed in Appendix D, the order-parameter correlation function is affected not only by phase/charge fluctuations but also by spin fluctuations. In fact, the identification of the $\{S(\mathbf{R}_t)\}$ variables as the chiral spin degrees of freedom discussed in the Introduction follows from their relation with the correlation function. In Appendix D we show that, in the low temperature regime,

$$g_{ij} \simeq \left\langle \exp\left(\frac{2\pi i}{3} S(\mathbf{R}_t)\right) \right\rangle \tag{2.14}$$

for a nearest-neighbor pair of sites $\langle ij \rangle$ that borders a triangular plaquette having integer index $S(\mathbf{R}_t)$. If $S(\mathbf{R}_t) = \pm 1$, then the phase-difference between two adjacent sites must equal $\pm 2\pi/3$ so we indeed can identify $S(\mathbf{R}_t)$ with the Ising variable discussed in the Introduction. By adding the phase-differences of two nearest neighbors, one finds that for two *next*-nearest neighbor sites that are part of the two adjacent triangular plaquettes $\mathbf{R}_t$ and $\mathbf{R}'_t$ which contain sites $i$ and $j$, (see Fig.6), one finds that the (low-temperature) correlation function for next-nearest-neighbor sites is given by:

$$g_{\langle\langle ij \rangle\rangle} \simeq \left\langle \exp\left(\frac{2\pi i}{3} \left(S(\mathbf{R}_t) + S(\mathbf{R}'_t)\right)\right) \right\rangle \tag{2.15}$$

We now can use Eq. (2.15) to include the *next*-nearest neighbor interaction term of the phase Hamiltonian that we ignored so far. Expanding the partition function to first order in the second neighbor interaction gives:

$$Z \cong Z_0 \left(1 - \beta_2 \operatorname{Re} \sum_{\langle\langle ij \rangle\rangle} g_{\langle\langle ij \rangle\rangle}\right) \tag{2.16}$$

where $\beta_2 = V_2 / k_B T$. Using Eq. (2.15) in Eq. (2.16), and exponentiating, one can include the second-neighbor interaction as a contribution to the spin Hamiltonian:

$$H_S \cong h\sum_{\mathbf{R}_t}\left(S(\mathbf{R}_t)+2f\right)^2 + J\sum_{\langle \mathbf{R}_t,\mathbf{R}'_t\rangle}\cos\left(\frac{2\pi}{3}\left(S(\mathbf{R}_t)+S(\mathbf{R}'_t)\right)\right) \qquad (2.17)$$

with $J = 2V_2$. Note that the first term of the spin Hamiltonian is minimized by choosing $S(\mathbf{R}_t)$ equal to $-2f$, so $f$ acts as a field conjugate to the chiral order parameter $\langle S(\mathbf{R}_t)\rangle$. If the two spin variables are opposite, then the second term equals $+J$. If the two spin variables are the same, then it equals $-J/2$ for $S = 1$ and $-J$ for $S = 3$. We thus can - crudely - interpret $J$ as a "ferromagnetic" exchange constant for chiral spin variables.

Finally, we show in Appendix D that if the two sites $\langle ij\rangle$ are on opposite sites of a *domain-wall* of the spin variable, then the domain wall produces loss of phase coherence. In effect, the spin-wave prediction $g_{SW} \propto \dfrac{1}{|\mathbf{R}_{ij}|^\eta}$ only applies for two sites that are part of the same domain of the spin degree of freedom. With Eqs (1.7), (1.10) and (1.17) we are now in a position to discuss the effect of chirality on the groundstate, low-temperature excitations, and phase-diagram of the system.

## III) Groundstate and Defect Structures.

In this Section we determine the different groundstates of the spin-charge Hamiltonian as a function of the chirality parameter $f$ and describe the corresponding counterion configurations of the Wigner crystal. Following this, we discuss defect structures of the groundstate, which will play an essential role when we construct the phase diagram.

### A) Groundstates

It follows from the phase Hamiltonian that the energy of a spin-charge structure is a periodic function of $f$ under the operation $f \to f + 3$ while under the operation $f \to -f$, the energy is invariant when the signs of all spin variables are inverted as well. In determining the groundstate structure one can thus be restricted to the interval $0 < f < 3/2$. The structure of the groundstate is obtained by minimizing the spin-charge Hamiltonian. The charge Hamiltonian is minimized by imposing $Q(\mathbf{R}_h) = 0$ on all hexagonal plaquettes and spin structure must be obtained by minimizing

$$H_S \cong h \sum_{\mathbf{R}_t} \left( S(\mathbf{R}_t) + 2f \right)^2 + J \sum_{\langle \mathbf{R}_t, \mathbf{R}'_t \rangle} \cos\left( \frac{2\pi}{3} \left( S(\mathbf{R}_t) + S(\mathbf{R}'_t) \right) \right)$$

subject to the constraint that for every hexagonal plaquette, the six spin variables surrounding the plaquette must obey the condition that $\frac{1}{6}\sum_{\mathbf{R}_t}' S(\mathbf{R}_t)$ is an integer.

### i) *Chiral Groundstate*

For $0 < f < 1$ the first term of the spin Hamiltonian is minimized if $S(\mathbf{R}_t) = -1$ on all triangular sites so $\frac{1}{6}\sum_{\mathbf{R}_t}' S(\mathbf{R}_t) = -1$. The corresponding counterion arrangement, shown in Fig.7A, is the helical structure discussed in the Introduction, with a phase-difference of $2\pi/3$ between adjacent counterion columns. The pitch of the counterion helix is fixed at $p_{el} = 2d$. The chiral groundstate has optimal stability at the center of the interval, i.e., $f = 1/2$, where the "field energy", first term of Eq. (2.17), is zero. In terms of the phase Hamiltonian Eq. (2.2), the optimal phase shift $2\pi/3$ equals at this point. This structure is known as a "$\mathbf{q} = 0$" state in the theory of the Kagomé XY anti-ferromagnet[16,17,18].

### ii) *Achiral Groundstate*

In the range $1 < f < 2$, the first term of the spin Hamiltonian is minimized by $S(\mathbf{R}_t) = -3$ on all triangular sites, with $\frac{1}{6}\sum_{\mathbf{R}_t}' S(\mathbf{R}_t) = -3$. The counterion arrangement, shown in Fig. 7C, is purely *achiral* in this case and composed of stacks of counterions perpendicular to the direction of the polymers. Note that $f = 3/2$, at the center of the achiral range, corresponds to the condition $A_{ij} = 2\pi f / 3 = \pi$. The phase Hamiltonian Eq. (2.2) reduces to the conventional *unfrustrated* XY model in this case.

iii) *Staggered Chiral Groundstate*

As shown in Fig.7B, the constraint condition also can be satisfied by assigning $S(\mathbf{R}_t) = -1$ and $S(\mathbf{R}_t) = -3$ to alternate triangular plaquettes surrounding a hexagonal plaquette, with $\frac{1}{6}\sum_{\mathbf{R}_t}' S(\mathbf{R}_t) = -2$. This state is, in fact, doubly-degenerate owing to the symmetry under interchanging the spins $S(\mathbf{R}_t) = -1$ and $S(\mathbf{R}_t) = -3$. This "staggered" phase can only be the groundstate in the regime $f \simeq 1$ where the chiral and achiral states are nearly degenerate and if the second neighbor interaction is included as well. The actual stability conditions are:

$$f > 1$$
$$\beta_2 > \frac{4\pi^2}{27}\beta_1(f-1) \tag{3.1}$$

This groundstate is only stable in a narrow region near $f \simeq 1$ since $V_2 \ll V_1$. In terms of the phase pattern, $\phi_i$ increases (or decreases) by $2\pi$ rather then $4\pi$ as the hexagonal plaquettes are circuited. The counterions wind around the polyelectrolyte cores with 1/2 the pitch of the uniform chiral groundstate.

An important result of this section is that, apart from this small interval around $f = 1$, the local constraints Eq. (2.10) stabilize either the chiral state with the "electrostatic" pitch, $p_{el}$ or the achiral state and appear to forbid any more complex chiral structures with a pitch intermediate between 0 and $p_{el}$. The results of the groundstate analysis are summarized in Figure 8.

B) Defects

The constraint conditions Eq.(2.10) also play an important role in determining the defect structures of the chiral groundstate. Because $Q(\mathbf{R}_h) - \frac{1}{6}\sum_{\mathbf{R}_t}' S(\mathbf{R}_t)$ has to be an integer, the topological charge $Q$ can only adopt the values $Q = \pm 1/3, \pm 2/3, \pm 1,..$ Another consequence of this constraint is that the excitations in general are expected to have a combined charge-spin structure [17].

## i) *Integer Charge Excitations.*

The integer charge excitations $Q(\mathbf{R}_h) = \pm 1$ of the $S(\mathbf{R}_t) = -1$ chiral groundstate are, exceptionally, *pure* charge defects that require no spin flips to satisfy the constraints. In terms of the counterion distribution, integer charge excitations represent *screw dislocations* in the Wigner crystalline groundstate with the phase $\phi$ changing by $\pm 2\pi$ in circuits enclosing charged hexagonal plaquettes. If the spin degrees of freedom are kept fixed in one of the two chiral groundstates, then the charge Hamiltonian $H_h$ can be used without further restrictions to obtain the energy of a distribution of integer-charged defects. Typical low-energy integer defect structures are then the usual $Q(\mathbf{R}_h) = \pm 1$ bound dipoles that characterize the low-temperature phase of the XY model.

In the limit $f = J = 0$, integer charge defects are however unstable because the chemical potential $\mu$ of *one* $Q(\mathbf{R}_h) = 1$ charge exceeds the sum of *three* chemical potentials equal to $\mu/9$, the chemical potential of a $Q(\mathbf{R}_h) = 1/3$ charge. Integer charges thus could break up into three fractional charges. A simple example is shown in Fig.9A: a $Q(\mathbf{R}_h) = 1$ charge is decomposed into three $Q(\mathbf{R}_h) = 1/3$ charges placed at the corners of an equilateral triangle of nine flipped spins surrounding the original hexagonal plaquette that carried the original integer charge. The spin flips are here required in order to satisfy our local constraints that $Q(\mathbf{R}_h) = 0$ except on the three vertices of the triangular domain.

Because of the logarithmic repulsion between the three fractional charges, the "electrostatic" energy cost of the triangle can be reduced by allowing the enclosed area $A$ of the triangle to grow. However, chirality and exchange interaction both act to stabilize integer defects against this form of decomposition. If a "field energy" cost $8hfA$ is imposed on the flipped spins then, by comparing the repulsive interaction potential $K \ln A^{1/2}$ between the fractional charges with the field energy term, it follows that the total energy is minimized for $A^* \sim \frac{K}{h} f^{-1}$. A similar argument shows that $A^* \sim (K/J)^2$ for the achiral case when $f = 0$. We conclude that integer defects of the chiral groundstate can be stable defects on larger length scales and that they can be endowed with an interior structure composed of fractional charges.

## ii) *Single Spin Flips.*

Figure 9B shows a single spin flip of the $S(\mathbf{R}_t) = -1$ groundstate. The constraints produce three $Q(\mathbf{R}_h) = 1/3$ charges on the three hexagonal plaquettes surrounding the triangular plaquette. This triplet of defects carries a net topological charge of $Q = 1$, so a single spin flip would carry a logarithmically divergent energy cost. This can be avoided by introducing a compensating $Q = -1$ integer charge, for example by replacing one of the three $Q(\mathbf{R}_h) = 1/3$ charges on the three hexagonal plaquettes with a $Q(\mathbf{R}_h) = -2/3$ charge.

### iii) *Chiral Droplets*

Straight chiral domain boundaries do not couple to the charge degree of freedom, but a $60^0$ or $120^0$ bend or kink of a domain boundary requires a fractional charge $Q(\mathbf{R}_h) = 1/3$ at the location of the kink. Closed domains of inverted spins - *chiral droplets* - can be constructed only by introducing fractional charges along the perimeter. Figure 10A shows such a domain in the form of a parallelogram, with straight boundaries along two of the crystallographic directions, with alternating $Q(\mathbf{R}_h) = \pm 1/3$ charges on its four corners (with net charge zero). The energy cost of a domain of size $\ell$ of this type can be estimated as:

$$E_{cd}(\ell) \simeq f^* \ell^2 + J \ell + \gamma \ln \ell \qquad (3.2)$$

with $f^* = 8hf$. The last term is the logarithmic Coulomb potential with $\gamma \propto K$. Note that it follows from Eq. (3.2) that the energy cost of separating fractional charges $Q(\mathbf{R}_h) = \pm 1/3$ charges by a distance $\ell$ scales either linearly with $\ell$ if $f^* = 0$ and quadratically otherwise.

Adding more fractional charges along the boundary allows for more complex perimeter shapes. Due to the attraction between charges of opposite sign, right- and lift-bending kinks attract and form bound dipole pairs[19]. A fractional dipole along an otherwise straight perimeter is equivalent to a *step,* or "jog", of the perimeter with a step size proportional to the spacing of the pair. The direction of the step is determined by the direction of the dipole. Domain boundaries of arbitrary shape can be constructed this way. An example of such a domain is shown in Figure 10B. A compact droplet of area $A$ with a perimeter of length $\ell \sim A^{1/2}$ that contains $n$ such steps would have an energy of order

$$E_{cd}(A, \ell, n) \simeq f^* A + J \ell + \varepsilon n \qquad (3.3)$$

Note that there are no longer a logarithmic term since all fractional charges come in dipole pairs. This type of chiral droplet domain is in fact charge effectively neutral.

### iv) *Chiral Worms.*

A second way of combining the fractional charges along a chiral domain boundary into dipole is by bringing the opposite sides of the domain together pairing fractional charges of opposite sign located on opposite sides. The opposing walls are effectively bound together, producing an extended *linear* structure. The bound pairs of $Q(\mathbf{R}_h) = 1/3$ fractional dipoles act here either as *endpoints* of a string of flipped spins along one of the crystallographic directions, or as $\pm 120^0$ *kinks* of the string, or as *branch-*

*points* (see Fig.11). Strings of flipped spins decorated with $Q(\mathbf{R}_h) = \pm 1/3$ dipoles will be referred to as *chiral worms*. The energy cost $E_{cw}$ of an (unbranched) chiral worm of length *m* with *n* kinks can be estimated as

$$E_{cw}(m,n) \simeq f^* m + \varepsilon(n+2) \tag{3.4}$$

with $f^* = 8hf$ and $\varepsilon \sim 2\mu/9$ (which is also the energy cost of a branch-point).

## IV) Phase Diagram

In the first part of this section, we will consider separately the statistical mechanics of the topological charges and the chiral droplets. In the second part, the results are combined to construct the phase diagram.

A) Statistical Mechanics of Spin and Charge Fluctuations

*i) Chiral Droplet Fluctuations.*
We have seen that by including fractional dipoles, we can construct chiral domains of arbitrary size and shape. The statistical mechanics of thermally excited chiral domains of this type will be treated by analogy with domain fluctuations of the 2D Ising model. In the Fisher droplet description, the critical properties of a 2D Ising model are obtained by summing over thermally excited domains of different size and shape [20]. Specifically, if we define $\Pi = F - MH$ to be the thermodynamic potential with $F$ the free energy per site, $M$ the magnetization per site and $H$ the applied field, then $\Pi$ (relative to the potential for uniformly aligned groundstate) can be expressed as the sum

$$\Pi / k_B T \propto \sum_{A,L} G(A,L) e^{-\beta \Delta E(A,L)} \tag{4.1}$$

Here, $G(A,L)$ is the number of configurations of an inverted droplet of area $A$ and perimeter length $L$ and with $\Delta E(A,L) \sim h^*A + J^*L$ the energy cost of the droplet ($J^*$ is the exchange constant). To analyze the dependence of $\Pi$ on temperature we consider the dominant contributions to the sum in Eq. (4.1) coming from asymptotically large domain fluctuations. For these large domains the dominant perimeter length of a compact droplet is estimated as $L(A) \sim \ell_0 A^{1/2}$ and one estimates $G$ by treating the perimeter as a closed, self-avoiding $L$-step walk:

$$G(L) \propto \frac{z^L}{L^{2\theta}} \tag{4.2}$$

Here, $z$ is a constant that depends on the geometry of the lattice and $\theta$ a critical exponent. The sum can then be written as

$$\Pi(T)/k_B T \propto \sum_A \left(e^{-\beta h^*}\right)^A \left(e^{-\beta \Sigma(T)}\right)^{A^{1/2}} / A^\theta \tag{4.3}$$

with $\Sigma(T) = J - k_B T \ln z$ an effective, temperature-dependent line tension. For $h^* = 0$, $\Pi(T)$ develops a power-law singularity at a critical temperature $k_B T_\chi = J^*/\ln z$ where the effective line tension $\Sigma(T)$ vanishes. The mean droplet size $\xi$ - the correlation length - diverges at this temperature as $\xi(\tau) \propto |\tau|^{-\nu}$ where $\tau = (T/T_\chi - 1)$ is the reduced temperature and $\nu$ the correlation length exponent. Below this critical temperature, $h^* = 0$ marks a line of first-order transitions.

We can apply the droplet description to the chiral domain fluctuations. However, due to the finite cost of a kink-pair step, chiral droplets must have an *elongated* shape and be oriented, on average, along one of the three axes of the Kagomé lattice. If a domain wall bordering such a droplet contains $L$ steps then it will have a length of order $L\ell_0$, with $\ell_0 \sim e^{\beta\varepsilon}$ the mean spacing between steps, and a width of order $L$. The droplet energy is then of the form $\Delta E(A,L) \sim h^*A + J^*L$ with $h^* = f\,h\ell_0$ and $J^* = J\ell_0$. Using this rescaling, the droplet energy, and hence the expression for the thermodynamic potential $\Pi$, is of the same form as Eq. (4.5). When $f = 0$, the mean size of the domains is thus infinite above a critical temperature $k_B T_\chi \approx J\ell_0$. Because $\ell_0 \sim e^{\beta\varepsilon}$ is temperature dependent, this is a self-consistency equation for the critical temperature with solution:

$$k_B T_\chi \sim \frac{V_1}{\ln(V_1/V_2)} \tag{4.4}$$

where we have assumed that $V_1 \gg V_2$ [21].

Fisher has showed that for non-zero applied fields [22], droplets are highly elongated having a long axis $\xi_\parallel(h^*, \tau)$ given by

$$\xi_\parallel^{-1}(h^*, \tau) - \xi^{-1}(\tau) \propto (h^*)^{2/3} \tag{4.5}$$

where $\tau = (T/T_\chi - 1)$. The width $\Delta x \sim (h^*)^{-1/3}$ of the stringy domain is small compared to the length, though it diverges for $h^* = 0$. We will assume the same scaling relation for the anisotropic chiral droplets apart from a rescaling of the correlation length along the long axis by a factor of $\ell_0$.

*ii) Integer and Fractional Charge Unbinding.*

The finite temperature statistical mechanics of the charge degrees of freedom will be described by the Kosterlitz-Thouless (KT) renormalization group (RG) equations. It is convenient, for the following, to express these KT equations in the "dielectric" language where thermally excited dipoles with a smaller spacing screen the interaction between pairs of topological charges with a larger spacing. This screening is described by a distance dependent dielectric constant $\varepsilon(r)$ for charges separated by a distance $r$. The force $F(r)$ between two integer charges of opposite sign, and their potential energy $V(r)$, are related to $\varepsilon(r)$ by

$$F(r) = -dV/dr = -\frac{KQ^2}{2\pi\varepsilon(r)r} \qquad (4.6)$$

At short distances $r$ comparable to the lattice constant $D$, Eq. (4.6) should reduce to the "bare" interaction of the charge Hamiltonian so the dielectric constant at short length scales would be $\varepsilon_0 = 1$. Let $n(r) = n_0 e^{-\beta V(r)}$ be the area density of dipoles with $n_0$ a normalization factor and define the related quantity $y_Q(r)$:

$$y_Q(r) \equiv n_0 r^2 e^{-\beta V(r)} \qquad (4.7)$$

The KT renormalization group equations can then be expressed as

$$\begin{aligned} r\frac{dy_Q}{dr} &= 2\left(1 - \frac{\beta KQ^2}{4\pi\varepsilon(r)}\right)y_Q \\ r\frac{d\varepsilon}{dr} &= \pi\beta KQ^2 y_Q^2 \end{aligned} \qquad (4.8)$$

The first equation is the derivative of Eq. (4.7) while the second equation gives the contribution to the renormalization of the dielectric constant from pairs of charges with spacing between $r$ and $r + dr$ [23]. The initial conditions are that $y_Q(r = D)$ is proportional to the "bare" activity of a charge and $\varepsilon(r = D) = 1$. The separatrix between flow lines that end at $y = 0$ and flow lines with divergent $y$ crosses the $y = 0$ line at the KT transition temperature $\beta_{KT} KQ^2 / 4\pi\varepsilon_0 = 1$. The $Q = 1$ KT unbinding temperature will be denoted by $T_1$. In principle, the unbinding temperature for fractional charges, $T_{1/3} \approx \frac{1}{9}T_1$ would be reduced by a factor $(1/3)^2$ compared with $T_1$ [17]. Recall however that the $Q = 1/3$ charges are confined as well by chiral domains, which acts as a separate constraint not included in the KT equations. In the next section, we combine the KT equations with the droplet description to include this constraint.

## B) Phase Diagram

In order to construct a phase diagram, we will combine, in a physically intuitive manner, the KT equations of integer and fractional charges with the Fisher droplet description. Three critical temperatures will play a role: the KT transition temperature for integer charge unbinding $k_B T_1 \sim V_1$, the KT transition temperature $T_{1/3}$ for fractional charge unbinding, which is small compared to $T_1$ $\left(T_{1/3} \sim \frac{1}{9} T_1\right)$ and the Ising transition temperature $k_B T_\chi \sim \frac{V_1}{\ln(V_1/V_2)}$. Since $V_2/V_1$ is small compared to one, we can assume that both $T_\chi$ and $T_{1/3}$ are small compared to $T_1$. We will focus first on the case $T_{1/3} < T_\chi < T_1$.

i) $T_{1/3} < T_\chi < T_1$

Recall that fractional charge unbinding demands the production of a chiral domain of at least the size of the separation between the charges. Since the size of chiral domains that are produced by thermal fluctuations is limited by the chiral correlation length $\xi(\tau)$, *we can treat $\xi(\tau)$ as a "finite size" limit on fractional charge unbinding at $T_{1/3}$*. The free energy cost for a fractional pair to be separated by a spacing $\ell$ large compared to $\xi(\tau)$ would scale as $\ell$ times the domain line tension $\Sigma(\tau)$ and thus rapidly prohibitively large compared to $k_B T$.

If the chiral correlation length acts as a finite size limit, then there can be no unbinding transition at $T_{1/3}$. Above $T_{1/3}$ fractional charges can only *partially* unbind, but they will remain confined inside chiral domains. The partial unbinding has however an important consequence, namely a sharp rise in the effective dielectric constant. This effect can be compared with the growth of the dielectric constant that is produced by imbedding a distribution of finite-sized conducting domains inside an insulating matrix. To estimate this growth of the dielectric constant starting near $T \sim T_{1/3}$, we can simply integrate the KT equations for $Q = 1/3$ up to the finite-size limit $r = \xi(\tau)$:

$$r \frac{dy}{dr} \sim 2\left(1 - \frac{T_{1/3}}{T}\right) y$$
$$r \frac{d\varepsilon}{dr} \sim 16\pi^2 \frac{T_{1/3}}{T} y^2$$
(4.9)

The solution is

$$y(\xi) \sim \xi^{2\left(1-\frac{T_{1/3}}{T}\right)}$$
$$\varepsilon(\xi) \sim \left(1-\frac{T_{1/3}}{T}\right)^{-1} \xi^{4\left(1-\frac{T_{1/3}}{T}\right)}$$
(4.10)

According to Eq. (4.10), the dielectric constant indeed grows rapidly above $T_{1/3}$ and, formally, diverges at the critical temperature $T = T_\chi$ of the droplet model. However, it follows from (4.8) that if the dielectric constant increases beyond the threshold $\varepsilon_1 = \beta K / 4\pi = T_1 / T$, then *integer*-charged screw dislocations - which are *not* confined to chiral domain fluctuations - can unbind. The increase of the dielectric constant due to the fractional charges thus leads to the melting of the chiral Wigner crystal at a temperature *below* $T_\chi$. The condition for melting by integer charge unbinding can be expressed as the condition $\varepsilon_1 = \varepsilon(\xi)$. This translates into the condition that in the Wigner crystalline phase, the chiral correlation length cannot exceed the limit:

$$\xi^* \sim \left(\frac{T_1}{T}\left(1-\frac{T_{1/3}}{T}\right)\right)^{\frac{1}{4(1-T_{1/3}/T)}}$$
(4.11)

Combining Eq. (4.11) with the relation $\xi_\parallel^{-1}(\tau, h^*) - \xi^{-1}(\tau) \propto (h^*)^{2/3}$ for the anomalous dependence of the correlation length of the 2D Ising model on field strength, we can compute the reduction $\Delta\tau$ of the melting transition temperature below $T_\chi$. If the chiral field is zero then the reduction is

$$\Delta\tau(0) \sim \left(\frac{T_\chi / T_1}{1 - T_{1/3}/T_\chi}\right)^{\frac{1}{4\nu(1-T_{1/3}/T_\chi)}}$$
(4.12)

Note that this reduction of the critical temperature becomes large when $T_{1/3}$ approaches $T_\chi$. In the presence of a field, the melting temperature increases with the field strength as:

$$T(h^*)/T_\chi - 1 \propto -\left(\Delta\tau(0)^\nu - c_0(h^*)^{2/3}\right)^{1/\nu}$$
(4.13)

The dependence of the melting temperature on the chiral field strength thus has a cusp with exponent 2/3 at $h^* = 0$. This sharp increase of the melting temperature as a function of the intrinsic chirality is due to the reduction of the size of the chiral domains by the applied field[24].

*ii)* $T_\chi < T_{1/3} < T_1$

The physics of the melting transition is altered when the unbinding temperature for fractional charges exceeds the critical temperature $T_\chi$. For $T > T_\chi$ and when $f = 0$, domain walls extend across the system. As discussed in Appendix D, the appearance of domain wall fluctuations of infinite size leads to exponential decay of the correlation function $g_{ij} = \langle e^{i(\phi_i - \phi_j)} \rangle$. It follows that, for $f = 0$, the chiral critical temperature $T_\chi$ coincides with the loss of periodic order along the *z* direction associated with counterion charge-density waves of wavevector $G = 2\pi/d$. However, as noted by Huse and Rutenberg for the analogous regime in Josephson Junction arrays [25], the correlation function $g_{ij}^3 = \langle e^{3i(\phi_i - \phi_j)} \rangle$ is *not* subject to the de-phasing by domain-wall boundaries intervening between *i* and *j* [25], since the phase jump across a boundary is a multiple of $2\pi/3$. This means that fractional charges do *not* unbind at $T_\chi$. Over the temperature interval $T_\chi < T < T_{1/3}$ and $f = 0$, $g_{ij}$ decays exponentially but $g_{ij}^3$ decays as a power-law. Correlation functions associated with all charge density modes with wavenuber *nG* decay exponentially, except for the modes where *n* is an integer multiple of 3 [14].

In this intermediate phase, the counterion density-density correlation function $\langle \delta\rho_i(0)\delta\rho_j(z)\rangle \sim \mathrm{Re}\left[e^{-i3Gz}g_{ij}^3\right]$ decays as a powerlaw. The counterions are restricted to parallel planes spaced by a distance $d/3$ perpendicular to the direction of the polymers. As in the chiral groundstate, the counterion columns have the usual $2\pi/3$ phase relation between adjacent columns. However, in the chiral *groundstate* the counterions of each layer occupy only one of the three possible sublattices, say A, B, and C, in alternating sequence from layer to layer. In the achiral phase, the counterions occupy these three sublattices with equal probability in a given plane. As this achiral phase maintains periodic order along the *z* direction (albeit of a shorter periodicity), the polyelectrolyte bundle continues to have a finite shear modulus.

Above $T_\chi$, but below $T_{1/3}$, integer and fractional defects both remain bound. Above $T_{1/3}$ the fractional dipole/kinks decorating chiral domain walls unbind leading to exponential decay of $g_{ij}^3$. This second transition completes the melting of the Wigner crystal as now correlations of all density modes *nG* are short-ranged. It is a challenging problem to determine whether the transition between the chiral and achiral Wigner crystals persists for finite values of *f*. Because *f* is a field conjugate to the chiral order parameter, it would seem that this transition could not extend to finite values of *f*. However, at lower temperatures ($T < T_\chi$) and finite *f*, the correlation function $g_{ij}$ of the primary order parameter exhibits the usual power-law decay whereas Appendix D provides arguments that for higher temperatures and finite *f*, $g_{ij}$ must decay exponentially, again due to dephasing by chiral domain walls that span the system. This would suggest that there can be a phase transition also at finite *f*, as characterized by the change of the correlation function of the primary order parameter. Because $f$ is finite, this would however not be a transition from a chiral and to a fully achiral state.

A conjectured phase diagram is shown in Fig.12. For the case of $T_{1/3} \gg T_\chi$, we expect the temperature at which $Q = 1/3$ charges unbind to have only a weak dependence on $f$ since at this temperature, the chiral field is too weak to restrain domain fluctuations. As $T_{1/3}$ approaches $T_\chi$ from above, the melting temperature of the Wigner crystal will develop a stronger dependence on $f$, as discussed above for the case where $T_{1/3} < T_\chi < T_1$.

## iii) Strong chirality

For larger values of the chirality parameter $f$, spin flips from $S = -1$ to $S = -3$ become competitive with spin flips from $S = -1$ to $S = +1$. The two spin-flip energies are in fact equal at $f = 1/2$ while for larger values of $f$, flips from $S = -1$ to $S = -3$ become increasingly favorable. The area density of fractional dipole pairs and the polarizability start to rise as $f$ increases beyond 1/2 so $f = 1/2$ must mark a maximum of the transition temperature $T(f)$. The transition temperature is expected to have a second minimum around $f = 1$. In this region, domain wall fluctuations involve walls separating chiral $S = -1$ regions and regions that are in the staggered chiral groundstate of Figure 7B. Domain boundaries separating the uniform $S = -1$ state and the staggered $S = -1 / S = -3$ state carry a finite density of fractional charges, *even when domain walls are straight*, so the bare line tension restraining these droplet fluctuations must be of the order of $V_1$. We thus expect the melting temperature at $f = 1$ to be higher than at $f = 0$. Similar arguments apply to the transition between the staggered state and the achiral $S = -3$ state. Finally, due to the symmetries under $f \to -f$ and $f \to f+3$, $f = 3/2$ must correspond to a local maximum in the melting temperature of the Wigner crystal.

# V) Conclusion

The central conclusion of this study is that chirality plays a fundamental role in the thermodynamics of counterion Wigner crystals inside biopolymer bundles. We found that the groundstate of the Wigner crystal exhibits *spontaneous chiral symmetry breaking*: even when the biopolymers themselves are not chiral, the counterion columns *still* arrange themselves in a pattern of helical spirals, with a phase shift of $2\pi/3$ between neighboring columns (the "$\mathbf{q}=0$" state). The origin of chiral symmetry breaking was found to lie in *frustration*: the hexagonal geometry of a biopolymer bundle does not allow counterion columns to minimize pairwise their mutual electrostatic repulsion, and the lowest-energy compromise structure is helical. The helical groundstate of the Wigner crystal is not altered when intrinsic chirality of modest strength is included, except that the sign of the helicity of the biopolymers determines the sign of the helicity of the counterion array.

We also found that, as the temperature is raised, the Wigner crystal melts by the unbinding of screw dislocations in the displacement pattern of the counterion columns along the direction of the polymers. The melting temperature $T(f)$ depends sensitively on the strength $f$ of the intrinsic chirality: it has a singular minimum at $f=0$, with an $f^{2/3}$ cusp, and the minimum melting temperature $T(f=0)$ is small compared to the maximum melting temperature $T(f=1/2)$. Chiral domains produced by thermal fluctuations play a key role in determining the melting temperature.

We found that the melting scenario takes a surprise turn if the electrostatic interaction between next-neighbour counterion columns becomes very weak, e.g., if the salt concentration of the solution is high: the chiral Wigner crystal first melts into an achiral Wigner crystal with a reduced periodicity along the $z$ direction, $d/3$. As the temperature is raised, the achiral crystal transforms into a phase with full translational symmetry through the unbinding of 1/3 *fractional topological defects*. The intermediate achiral phase disappears as the range and strength of the electrostatic interaction increases. It is interesting to note a curious similarity of this melting scenario with the well-known Halperin-Nelson-Kosterlitz-Thouless-Young two-stage melting scenario of 2D crystals, where *dislocation* unbinding first produces a hexatic mesophase which then transforms into an isotropic fluid by *disclination* unbinding[26]. The difference is that a dislocation with a large Burgers vector can be viewed as a pair 5-7 of disclinations while a screw-dislocation line of the Wigner crystal must be viewed as an assembly of three fractional 1/3 charges.

The model that we used in this paper to arrive at these conclusions has a number of important limitations. First, a given counterion column was described by a single phase variable. If the polyelectrolyte molecules are sufficiently long than we have to allow the phase to vary along the $z$ direction, i.e., we have to allow for compressional fluctuations of the counterion columns. In Appendix A, we show that longitudinal fluctuations indeed can be neglected for sufficiently short biopolymers but for longer chains, longitudinal de-phasing would turn the $D = 2$ statistical mechanics model treated this paper into a $D = 3$ problem. We believe that the general structure of the 3D phase diagram will be similar to that of Fig.4, but there is reason to be cautious. One might expect, from the results of this paper, that the 3D melting thermodynamics should be dominated by screw dislocation unbinding, which should be in the universality class of the 3D XY model. In a recent letter we investigated the effects of longitudinal phase fluctuations in our model for a simple special case, namely with the counterion columns restricted to a - still frustrated - hexagonal lattice rather than the Kagomé lattice[11]. Interestingly, the finite temperature classical statistical mechanics of Wigner crystal melting could be rigorously mapped onto the $T = 0$ quantum mechanics of the metal-insulator transition of a hexagonal Josephson-junction array in a magnetic field, i.e., with thermal fluctuations replaced by quantum fluctuations. Bending fluctuations of the polyelectrolyte chains played the role of gauge fluctuations of the vector potential. From the - highly non-trivial - critical properties of the corresponding quantum model[27], we were able to obtain the critical properties of the melting transition and they were *not* those of the 3D XY model, We thus must expect also for the present case that in the 3D limit the thermodynamics of the melting transition will not be in the universality class of the 3D XY model.

Could the same mapping be applied to include longitudinal phase fluctuations in the Kagome lattice model? The $f = 0$ line of the model indeed can be mapped onto a frustrated Josephson-junction array on a Kagomé lattice in a magnetic field. However, the $A_{ij}$ term of the phase Hamiltonian corresponds to a *staggered* magnetic flux. The quantum mechanics of metal-insulator transitions of Josephson-junction arrays with staggered flux Hamiltonians has not yet been investigated, either experimentally or theoretically, though staggered flux Hamiltonians *have* been studied in the context of models of high $T_c$ superconductivity, which - interestingly - also feature fractional charges ("anyons")[28].

A second important limitation of our model is that we restricted the counterions of a biopolymer to the contact lines between adjacent biopolymers. This would seem to be a reasonable assumption for counterions with high valency as the phonon mode for in-plane counterion displacements is necessarily coupled to that for the biopolymer displacement through the rotational invariance of the bundle. But given the high degree of frustration of the Wigner crystal ground state, the possibility of low-energy, lateral-displacements "defects" cannot be ruled out. To address both limitations, and also to search for possible Wigner crystal mesophases, we are currently carrying out numerical simulations on the polyelectrolyte bundles in the presence of polyvalent counterions[29].

## Acknowledgements

We would like to acknowledge the NSF for support under DMR Grant 04-04507 as well as the Kavli Institute for Theoretical Physics were some of this work was done. We also would also like to thank Joseph Rudnick and Subir Sachdev for helpful discussions.

## Appendix A: Debye-Hückel Theory and Model Parameters

In this section, we provide explicit expressions for the model parameters of the phase Hamiltonian on physical parameters, assuming the Debye-Hückel (DH) limit of aqueous electrostatics. In DH theory, the electrostatic interaction between two polyvalent point-like counterions of valency $Z$ is proportional to $V(r) \sim (Ze)^2 e^{-\kappa r}/r$ where the Debye parameter $\kappa$ is determined by the concentration of monovalent salt ions. The electrostatic interaction between the counterion columns $i$ and $j$ is of the form:

$$H_{ij}^{el} = \frac{e^2}{2} \int_0^L dz \int_0^L dz' \, \delta\rho_i(z) V_{ij}^{el}(z-z') \delta\rho_j(z') \tag{A.1}$$

assuming a counterion charge density of the form of Eq.2.1 and

$$V_{ij}^{el}(z-z') = V\left(\sqrt{D_{ij}^2 + (z-z')^2}\right) \tag{A.2}$$

Here, $D_{ij}$ is the in-plane seperation between columns at Kagomé sites $i$ and $j$. Using Eq. (2.1), we find the following interaction between the two columns,

$$H_{ij}^{el} = \sum_{n=1}^{\infty} V_{ij}^{(n)} \cos(n[\phi_i - \phi_j]) \tag{A.3}$$

where

$$V_{ij}^{(n)} = \frac{(Ze)^2 L}{4d^2} K_0\left(D_{ij}\sqrt{(nG)^2 + \kappa^2}\right) \tag{A.4}$$

and $K_0(x)$ a modified Bessel function. Note that $V_{ij}^{(n)}$ drops exponentially as a function of $n$. Retaining only the $n = 1$ terms yields the effective anti-ferromagnetic terms (with $A_{ij} = 0$) in Eq. (2.2).

Compressional modes of the counterions can be included by allowing variation of counterion phase variable $\phi_i$ along the $z$ direction. The inter-ion separation along the chain changes by $d \to d(1 + G^{-1} d\phi_i/dz)$. The associated energy cost is of the form:

$$H_C = \frac{C}{2}\int_0^L dz \left(\frac{d\phi_i}{dz}\right)^2 \qquad (A.6)$$

with C proportional to the compressional modulus. In DH theory the electrostatic self-energy per unit length of the counterion column would be given by,

$$U_{self}(d) = -\frac{(Ze)^2}{d^2}\ln\left(1-e^{-\kappa d}\right) \qquad (A.5)$$

It follows from Eq. A5 that the magnitude of the compressive modulus is of the form $C \sim (Ze)^2 e^{-\kappa d}$. The correlation function for phase fluctuations *along* the column is of the form:

$$\langle \delta\rho_i(0)\delta\rho_i(z)\rangle \sim \cos(Gz)\langle e^{i[\phi_i(0)-\phi_i(z)]}\rangle \qquad (A.7)$$

Neglecting inter-column coupling, one find that $\langle e^{i[\phi_i(0)-\phi_i(z)]}\rangle \sim e^{-|z|/\xi_z}$, with $\xi_z = 2C/k_B T$, which is of the order of $Z^2 \ell_B \ln(1/\kappa d)$, with $\ell_B$ the Bjerrum length. The theory presented in this paper, which neglects longitudinal phase fluctuations, thus is valid provide the chain length $L$ is less than $\xi_z$, which requires large Z and low salt concentration.

## Appendix B: Chiral Phase Shift

In this Appendix we present a simple microscopic model for computing the gauge-like term $A_{ij}$ appearing in Eq. (2.2). Consider the phases of two adjacent counterion columns, $\phi_1$ and $\phi_2$, as part of a hexagonal plaquette surrounding a polyelectrolyte molecule. The non-uniform charge distribution along the molecule produces a helical contribution to the electrostatic potential $\psi_1$, which has a periodicity $p$ along the direction of the polymer backbone. The electrostatic potential $U(z)$ along column 1 produced by the helical charge distribution is periodic with period p so $U(z) = U(z+p)$. The potential encountered by counterion column 2 is then the same function $U(z)$ but shifted along the z direction by a distance $\pm p/6$ due to the rotation around the charge spiral (the sign is determined by the handedness of the molecule). In a simple model, the electrostatic potential can be represented by just a single Fourier mode:

$$U(z) = U_0 \cos\left(G'[z+u(z)]\right) \qquad (B.1)$$

where $G' = 2\pi / p$ while $u(z)$ represents the local displacement of the *polyelectrolyte backbone* along the z direction, which is assumed to be compressible. If $\rho_G$ is the dominant amplitude of the counterion charge density modulation with wave number $G = 2\pi / d$, then the helical potential adds the following contribution to the phase Hamiltonian:

$$H_{12}^* = \rho_G U_0 \int_0^L dz \left\{ \cos(G'[z + u(z)]) \cos(Gz + \phi_1) + \cos(G'[z + u(z)] \pm p/6) \cos(Gz + \phi_2) \right\} \quad (B.2)$$

For slow variation of $u(z)$ along the backbone, this integral is dominated by the following term,

$$H_{12}^* \simeq \rho_G U_0 \int_0^L dz \cos\left( \Delta G z + \frac{1}{2}(\phi_1 + \phi_2) + G'u(z) \pm \frac{\pi}{6} \right) \cos\left( \frac{1}{2}(\phi_1 - \phi_2) \mp \frac{\pi}{6} \right) \quad (B.3)$$

where $\Delta G = G - G'$. This chiral energy is minimized for $du(z)/dz = -(\Delta G / G')$ with the counterion density modulation locked to the charge modulation of the polyelectrolyte, but this is resisted by the rigidity of the polyelectrolyte. If we define a one-dimensional compressional modulus B for the polyelectrolyte, then the combined energy has the form:

$$H_{12} = \int_0^L dz \left\{ B\left(\frac{du}{dz}\right)^2 + \Gamma_{12} \cos(\Delta G z + G'u(z) + \phi^*) \right\} \quad (B.4)$$

Here, $\Gamma_{12} = \rho_G U_0 \cos((\phi_1 - \phi_2)/2 \mp \pi/6)$ and $\phi^* = (\phi_1 + \phi_2)/2 \pm \pi/6$. This energy expression was minimized by Frank and van der Merwe in their study of misfit locations[30]. It follows from their results that if $(\Delta G / G')^2 > \Gamma_{12} / B$ then $u(z)$ locks into the minimum of the potential, i.e. $G'u(z) + \Delta G z + \phi^*$ is set to an odd or even multiple of $\pi$, depending on the sign of $\Gamma_{12}$. In this locked state, the energy depends on the phase difference as

$$\Delta H_{12} / L = -\left| \rho_G U_0 \cos((\phi_1 - \phi_2)/2 \mp \pi/6) \right| \quad (B.5)$$

The optimal phase difference of $\pm \pi / 3$ in Eq. (B.5) corresponds to a gauge term of magnitude $|A_{ij}| = \pi / 3$, i.e $f = \pm 1/2$, but note that this chiral term still must be combined with the direct nearest-neighbor electrostatic interaction.

The opposite regime, where the molecular compressibility resists lock-in, corresponds to an incommensurate "floating phase". Thermal fluctuations of $u(z)$ play an important role in this phase. We will expand the partition function in powers of $\Gamma_{12}$ and integrate the terms over $u(z)$. To lowest order in perturbation theory, one finds the following chiral contribution to the Hamiltonian,

$$\Delta H_{12} \simeq -\frac{|\Gamma_{12}|^2}{2} \int_0^L dz \int_0^L dz' \langle \cos(\Delta Gz + G'u(z))\cos(\Delta Gz' + G'u(z'))\rangle_0 \quad (B.6)$$

where $\langle \cdot \rangle_0$ denotes a thermal average with respect to the Boltzmann weight $\exp\left(-\beta B \int_0^L dz (du/dz)^2\right)$. Computing this expectation value one obtains an expression of the form:

$$\Delta H_{12} \simeq -V^* \cos\left(\phi_1 - \phi_2 \mp \frac{\pi}{3}\right) + C_0 \quad (B.7)$$

with

$$V^* = \frac{2\xi_z L |\rho_G U_0|^2}{1 + (\Delta G \xi_z)^2} \quad (B.8)$$

and

$$\xi_z = \frac{2B}{(G')^2 k_B T} \quad (B.9)$$

Note that the optimal phase shift in the floating phase again equals $\pm \pi/3$. Combining the chiral interaction with the "direct" anti-ferromagnetic interaction, $V_1^0 \cos(\phi_1 - \phi_2)$ gives an effective nearest-neighbor column interaction of the form of Eq. (2.2) with

$$V_1 = \sqrt{(V_1^0)^2 \mp \sqrt{3} V_1^0 V^* / 2} \quad (B.10)$$

and

$$\tan\left(\frac{2\pi f}{3}\right) = \mp \frac{\sqrt{3} V^*}{2V_1^0 - V^*} \quad (B.11)$$

Note that as $U_0 \to 0$ we return to the achiral limit, $f \to 0$. If the longitudinal phase fluctuations of Appendix B are included, then similar results are obtained, even if the backbone is incompressible, with the longitudinal counterion compressibility $C$ replacing $B$.

## Appendix C: Poisson Summation and Dual Model

In this appendix we provide details regarding derivation of the dual Hamiltonian, $H_d$, from the partition function in Eq. (2.5). Using the Poisson summation formula, the unconstrained sums in $Z_0$ over $j(\mathbf{R}_h)$ and $j(\mathbf{R}_t)$ can be written as integrals over the continuous fields $\Phi(\mathbf{R}_h)$ and $\Phi(\mathbf{R}_t)$ along with a discrete sum over integers all $m(\mathbf{R}_h)$ and $m(\mathbf{R}_t)$,

$$Z_0 = \sum_{\{m(\mathbf{R}_h),m(\mathbf{R}_t)\}} \int [d\Phi(\mathbf{R}_h)][d\Phi(\mathbf{R}_t)] \left( \prod_{\mathbf{R}_h} e^{2\pi i \Phi(\mathbf{R}_h)[m(\mathbf{R}_h)-3-2f]} \right) \left( \prod_{\mathbf{R}_t} e^{-S(\mathbf{R}_t)} \right) \quad \text{(C.1)}$$

with

$$S(\mathbf{R}_t) = \sum_{\mathbf{R}_h}' \frac{\beta_1^{-1}}{2} [\Phi(\mathbf{R}_h) - \Phi(\mathbf{R}_t)]^2 - 2\pi i \Phi(\mathbf{R}_t) \left( m(\mathbf{R}_t) + \frac{3}{2} + f \right) \quad \text{(C.2)}$$

where the prime on the sum indicates a summation over the three hexagonal sites $\mathbf{R}_h$ which border the triangular site at $\mathbf{R}_t$. This "action" is harmonic in $\Phi(\mathbf{R}_t)$, and these fields can be integrated our on each $\mathbf{R}_t$ to yield the following partition function,

$$Z_0 = \sum_{\substack{M(\mathbf{R}_h)=0,\pm1,\pm2,\ldots \\ S(\mathbf{R}_t)=\pm1,\pm3,\ldots}} e^{-\frac{\pi^2 \beta_1}{6} \sum_{\mathbf{R}_t}(S(\mathbf{R}_t)+2f)^2}$$

$$\times \int [d\Phi(\mathbf{R}_h)] \exp\left\{ -\frac{\beta_1^{-1}}{3} \sum_{\langle \mathbf{R}_h, \mathbf{R}_h' \rangle} [\Phi(\mathbf{R}_h) - \Phi(\mathbf{R}_h')]^2 + 2\pi i \sum_{\mathbf{R}_h} \Phi(\mathbf{R}_h) Q(\mathbf{R}_h) \right\} \quad \text{(C.3)}$$

where, after redefining the integer sums,

$$S(\mathbf{R}_t) = 3 - 2m(\mathbf{R}_t) \quad \text{(C.4)}$$

and

$$M(\mathbf{R}_h) = m(\mathbf{R}_h) - 3 \quad \text{(C.5)}$$

we arrive at the definition of defect "charge" in Eq. (2.10).

The continuous field $\Phi(\mathbf{R}_h)$ (defined on the hexagonal lattice connecting hexagonal dual sites, $\mathbf{R}_h$) plays the roles of an "electrostrostatic" potential which mediates interactions between "charges", $Q(\mathbf{R}_h)$. By integrating over $\Phi(\mathbf{R}_h)$ above, it can be shown that charges on sites $\mathbf{R}_h$ and $\mathbf{R}'_h$ interact via a potential $3\pi^2 \beta_1 G(\mathbf{R}_h, \mathbf{R}'_h)$ where $G^{-1}(\mathbf{R}_h, \mathbf{R}'_h)$ is the Laplacian operator define on the hexagonal lattice,

$$G^{-1}(\mathbf{R}_h, \mathbf{R}'_h) = 6\delta_{\mathbf{R}_h, \mathbf{R}'_h} - \delta_{\langle \mathbf{R}_h, \mathbf{R}'_h \rangle} \qquad (C.6)$$

Because $G(0)$ diverges as $\ln(R/D)$, where $R$ is the system size, it is useful to subtract this divergence from the interaction, $G'(\mathbf{R}_h) = G(\mathbf{R}_h) - G(0)$. The remaining portion is well approximated by,

$$G'(\mathbf{R}_h, \mathbf{R}'_h) \simeq -\frac{1}{2\sqrt{3}} \ln\left(\left|\mathbf{R}_h - \mathbf{R}'_h\right|/D\right) - \frac{1}{6} \qquad (C.7)$$

which holds to better than 1% for all $\mathbf{R}_h \neq \mathbf{R}'_h$. Upon inserting this form of the charge interaction into the partition function, we arrive at Eqs. (2.6) - (2.8). Note that the global defect charge neutrality condition is a direct result of the logarithmic divergence of $G(0)$ [31].

Appendix D: Fluctuations and Correlations in the Dual Model

In this appendix we describing how phase correlations in the original model of Eq. (2.2) are related to the statistical mechanics of the dual theory. The phase-phase correlation function is defined as:

$$g_{ij} = \left\langle e^{i(\phi_i - \phi_j)} \right\rangle \qquad (D.1)$$

At low temperatures the phase of $g_{ij}$ encodes the relative displacement along the $z$ direction of counterion columns at Kagomé sites $i$ and $j$. The counterion density-density correlation function is related to $g_{ij}$ by $\langle \delta\rho_i(0)\delta\rho_j(z)\rangle \sim \text{Re}\left[e^{-iGz} g_{ij}\right]$ so loss of phase coherence by transverse phase fluctuations implies a loss of position correlations of counterions in Wigner crystal.

We can compute $g_{ij}$ by noting that in the calculation of the average of $e^{i(\phi_i - \phi_j)}$ the "current conservation" condition on sites $i$ and $j$ is altered. To compute $g_{ij}$ we must instead sum over bond-integer configurations which satisfy,

$$\sum_k n_{ik} = -\sum_k n_{jk} = 1 \qquad (D.2)$$

and satisfy $\sum_k n_{mk} = 0$ for all other sites. This is most easily accomplished by summing over the circulating currents $j(\mathbf{R}_h)$ and $j(\mathbf{R}_t)$ in the presence of an *external current* passing from $i$ and $j$ [23] (see Figure 13). We denote the path of this external current by $\mathbf{r}_{ext}$, and note that the path may be chosen arbitrarily. The summation over $j(\mathbf{R}_h)$ and $j(\mathbf{R}_t)$ in the presence of the external current is conveniently accomplished by replacing $j(\mathbf{R}_h)$ with $j(\mathbf{R}_h) + \zeta(\mathbf{R}_h)$ in Eq. (2.5). Here, $\zeta(\mathbf{R}_h)$ is a function that is zero if $\mathbf{R}_h \notin \mathbf{r}_{ext}$ (meaning $\mathbf{r}_{ext}$ does not pass through a side of the hexagonal cell at $\mathbf{R}_h$), while if $\mathbf{R}_h \in \mathbf{r}_{ext}$, then $\zeta(\mathbf{R}_h) = \pm 1$. The sign is determined by the relative orientation of the external current and the circulating current $j(\mathbf{R}_h)$. Following this replacement, $\Phi(\mathbf{R}_t)$ and $\Phi(\mathbf{R}_h)$ can be integrated over, as in Appendix C, to show that,

$$g_{ij} = g_{SW} \langle g_S g_Q \rangle_d \tag{D.3}$$

Here, $g_{ij}$ is factored according to the three types of fluctuations that affect phase correlations: spin-wave - or phonon - fluctuations,

$$g_{SW} = \exp\left\{-\frac{\beta_1^{-1}}{12} \sum_{\mathbf{R}_h, \mathbf{R}_h'} \zeta'(\mathbf{R}_h) G(\mathbf{R}_h, \mathbf{R}_h') \zeta'(\mathbf{R}_h') + \frac{\beta_1^{-1}}{3} \sum_{\mathbf{R}_t \in \mathbf{r}_{ext}} \zeta^2(\mathbf{R}_t)\right\} \tag{D.4}$$

chirality fluctuations,

$$g_S = \exp\left\{\frac{2\pi i}{3} \sum_{\mathbf{R}_t \in \mathbf{r}_{ext}} \zeta(\mathbf{R}_t) S(\mathbf{R}_t)\right\} \tag{D.5}$$

and "charge" vortex fluctuations,

$$g_Q = \exp\left\{-i\pi \sum_{\mathbf{R}_h, \mathbf{R}_h'} \zeta'(\mathbf{R}_h) G(\mathbf{R}_h, \mathbf{R}_h') Q(\mathbf{R}_h')\right\} \tag{D.6}$$

The brackets $\langle \cdot \rangle_d$ in Eq. (D.3) indicate a thermal average with respect to the dual statistics in Eq. (2.6). The kernel, $G(\mathbf{R}_h, \mathbf{R}_h')$, is given is Eq. (C.7). The function $\zeta'(\mathbf{R}_h)$ is only non-zero for hexagonal cells with vertices and edges passed through by $\mathbf{r}_{ext}$. For each bond along the path $\mathbf{r}_{ext}$, $\zeta'(\mathbf{R}_h)$ receives a contribution $+2\zeta(\mathbf{R}_h)$ on the hexagonal cell $\mathbf{R}_h$ sharing an edge with that bond, and $\zeta'(\mathbf{R}_h)$ receives a contribution $-\zeta(\mathbf{R}_h)$ on the hexagonal sites which only share vertices with that particular bond. An example distribution of $\zeta''(\mathbf{R}_h)$ is shown in Figure 13. As noted in ref. 20, $\zeta''(\mathbf{R}_h)$ generates discrete along the path, $\mathbf{r}_{ext}$.

First, we consider the case of $i$ and $j$ which are nearest-neighbor sites. In this case $g_{ij}$ encodes the groundstate structure. For short separations spin-wave fluctuations are negligible and $g_{SW} \simeq 1$. Consider a nearest neighbor pair that forms the edge of the triangular plaquette at $\mathbf{R}_t$. The presence of non-zero defect charge on hexagonal sites leads to the following contribution to $g_Q$,

$$g_Q \simeq \exp\left\{-\frac{i}{2}\sum_{\mathbf{R}_h}\frac{(\mathbf{R}_h - \mathbf{R}_t)\cdot \hat{y}}{|\mathbf{R}_h - \mathbf{R}_t|^2}Q(\mathbf{R}_h)\right\} \tag{D.7}$$

where we have assumed that $|\mathbf{R}_h - \mathbf{R}_t| \gg D$. From (D.7) we see that the variables $Q(\mathbf{R}_h)$ describes vortex, or screw defect, phase patterns. While at high temperature a non-zero concentration of free vortices destroys phase coherence, at low temperatures vortices are confined to dipole pairs and the their effect on $g_Q$ is limited. In this regime, $g_{ij} \simeq \langle g_S \rangle_d = \langle e^{2\pi i \zeta(\mathbf{R}_t)S(\mathbf{R}_t)/3} \rangle_d$. It follows that the spin variables $S(\mathbf{R}_t)$ describe precisely the right- or left-handed helical arrangements of counterions around the triangular plaquette at $\mathbf{R}_t$ shown in Figure 2C.

In the limit that $|\mathbf{R}_{ij}| \gg D$ —where $\mathbf{R}_{ij}$ is the seperation between sites $i$ and $j$—it can be shown using the methods outlined in ref. 23 that spin-wave fluctuations lead to the following asymptotic contribution to $g_{ij}$,

$$g_{SW} \simeq e^{-\beta_1^{-1}/3}\left(\frac{|\mathbf{R}_{ij}|}{D}\right)^{-\eta} \tag{D.8}$$

with

$$\eta = \frac{1}{\sqrt{3\pi\beta_1}} \tag{D.9}$$

The algebraic decay of phase correlations in a two-dimensional XY system is a well-known consequence of spin-wave fluctuations [31]. It can also be shown that in the limit of large separations the contribution to phase correlations for vortex fluctuations has the same form as the unfrustrated XY model on the square lattice[23], namely,

$$g_Q \simeq \exp\left\{i\sum_{\mathbf{R}_h} Q(\mathbf{R}_h)\left[\theta(\mathbf{R}_h - \mathbf{R}_{ij}) - \theta(\mathbf{R}_h)\right]\right\} \tag{D.10}$$

where $\tan\theta(\mathbf{R}) = x/y$ and for simplicity the origin is set to site $i$.

The most significant difference between the long-distance properties of phase correlations of this model and simpler XY systems is due the presence of Ising-like fluctuations of chirality. In the $f = 0$ and $\beta_2 = 0$ limit of the Kagomé anti-ferromagnet, Cherapanov *et al.* demonstrated that states with $S(\mathbf{R}_t) = \pm 1$ occur with equal probability[14]. Therefore, even in the low-temperature limit where we are essentially restricted to $Q(\mathbf{R}_h) = 0$ groundstates the thermal average of $e^{2\pi i \zeta(\mathbf{R}_t)S(\mathbf{R}_t)/3}$ entering $g_S$ in Eq. (D.5) will be equal to $-1/2$ for each $\mathbf{R}_t \in \mathbf{r}_{ext}$. And hence for the achiral limit fluctuations of the chiral variables on the triangular sites lead to an exponential decay of correlations with distance,

$$g_{ij} \sim g_{SW}\langle g_S\rangle_d \sim \left(\frac{|\mathbf{R}_{ij}|}{D}\right)^{-\eta}(-2)^{-|\mathbf{R}_{ij}|/D}, \quad \text{for } f = \beta_2 = 0 \tag{D.11}$$

The above argument relies however on the presence of configurations with chiral domain walls of infinite size. Consider the case, with $f > 0$, where $S(\mathbf{R}_t)$ is - on average - aligned to the –1 state and where chiral domains with $S(\mathbf{R}_t)$ aligned to the +1 state are finite and charge-neutral. If neither site i are nor site j are inside one of the finite-sized minority domains, then, in computing $g_{ij}$, one can always find a path from i to j that avoids the minority domains. If the domains are indeed charge neutral then decay of $g_{ij}$ can only be due to spin-wave fluctuations so $g_{ij} \sim g_{SW}$, leading to algebraic decay with distance.

**Figure Captions**

Figure 1. Groundstate of polyvalent counterions (valency Z), shown as small spheres, neutralizing a single cylindrical macroion of diameter $\sigma_c$ (Fig.1a) or two cylindrical macroions in close proximity (Fig.1b). The counterions are excluded from the macroion interior volume. The charge per unit length of the macroion is $-e\rho_0$. In Fig.1a, counterions are positioned along a helix with a pitch that depends on the range of the electrostatic interaction. The axial spacing $Z/\rho_0$ between adjacent counterions is determined by the condition of charge neutrality. In Fig.1b, counterions are positioned along a column on the line of closest approach between the two macroions. The counterions produce a net short-range attraction between the macroions.

Figure 2. Groundstate of polyvalent counterions neutralizing an extended, hexagonal bundle of macroions. Fig. 2a: top view of the bundle. Counterions are positioned along columns situated on the sites of a Kagomé lattice. The columns define triangular "plaquettes" as indicated by a circle. Fig.2b: side of a prismatic triangular plaquette. Triangular plaquettes are frustrated because geometry prevents perfect pair-wise stagger of the three columns with their respective nearest-neighbors, as favored by electrostatic repulsion. Fig.2c: Minimum-energy compromise structure. Neighboring columns are shifted vertically by $d/3$. The counterions form a helical spiral. The sign of the helicity is not determined.

Figure 3. Schematic hase diagram. The horizontal axis $f$ is a measure of the intrinsic chirality of the biopolymers. The Wigner crystal can adopt three different structures. In the "chiral" state, counterions adopt the helical structure of Fig.2. At zero temperature, all triangular plaquettes have the same helicity sign determined by the sign of $f$. In the "staggered" chiral phase, the pitch of the counterion helix has doubled, while the counterions are arranged in identical layers in the a-chiral uniform phase (see Fig.7). Transitions between the different structures are first order, with spontaneous chiral symmetry breaking along the line $f = 0$. The solid lines indicate continuous melting transitions of the Wigner crystal by the unbinding of integer-screw dislocations (see Sec. IV). The phase diagram is periodic under $f \rightarrow f + 3$.

Figure 4. Helical charge distributions of the polyelectrolyte core (solid line) compared with the charge distribution of the counterions (Fig.4a). The pitch $p$ of the polyelectrolyte charge distribution in general need not equal the pitch of the counterion helix but "lock-in" will take place (white arrows) if the electrostatic interaction between the counterions and the helical charge distribution is strong. Fig.4 b shows the orientation of the terms $A_{ij}$ entering the phase Hamiltonian Eq. (2.1).

Figure 5. 5a: The Kagomé lattice (solid dots) and nearest-neighbor bonds of the Kagomé lattice (solid lines). The dual sites of the Kagomé lattice are shown as gray dots and the bonds of the dual lattice are shown as dashed gray lines. Fig.5b: current configurations on the bonds of the dual lattice that satisfy the conservation condition are constructed from currents circulating around the hexagonal and triangular plaquettes.

Figure 6. The "external" current that must be included to compute the contribution to the dual theory from the next-nearest neighbor interactions between $\phi_i$ and $\phi_j$ in Eq. (2.16). This term leads to a coupling between the spin variables $S(\mathbf{R}_t)$ located on the two shaded triangular sites (see Appendix D).

Figure 7. The three Wigner crystal groundstates (lower part) are shown with the corresponding chiral spin configurations (top part). The values of the integer indices $M(\mathbf{R}_h)$ are given inside the hexagonal cells and the values of $S(\mathbf{R}_t)$ are given inside the triangular cells. The topological charge $Q(\mathbf{R}_h) = M(\mathbf{R}_h) + \frac{1}{6}\sum_{\mathbf{R}_t}' S(\mathbf{R}_t)$ on the hexagonal sites is zero in all three cases. The stability range of the three states is indicated.

Figure 8. The groundstate structure of the chiral as functions of the ratio $V_2/V_1$ of the strength of next-nearest-neighbor and nearest-neighbor electrostatic interaction between counterion columns and the chiral parameter, $f$. Shaded regions correspond to chiral structures.

Figure 9. Composite spin-charge defects. Fig.9a: $Q = 1$ defect composed of three separate $Q(\mathbf{R}_h) = +1/3$ defects located on the vertices of an equilateral triangle. Inside the triangular domain, the spin variables are $S(\mathbf{R}_t) = +1$ (shaded), outside the domain $S(\mathbf{R}_t) = -1$ (white). Fig.9b: a charge-neutral single flip (shaded triangle) requires three fractionally-charged defects.

Figure 10. Chiral Droplets. Fig.10a shows a neutral chiral droplet in the form of a paralellogram with the minimum number of fractional defects decorating the four corners. $Q(\mathbf{R}_h) = +1/3$ charges are indicated as solid dots and $Q(\mathbf{R}_h) = -1/3$ as open dots. Note that the fractional charges are separated on length scales of the order of the droplet size. Fig.10b shows a more complex domain shape. The kink-steps of the domain wall are decorated by a distribution of $Q(\mathbf{R}_h) = \pm 1/3$ neutral dipole pairs. Here, $Q(\mathbf{R}_h) = +1/3$ charges are labeled as filled circles and $Q(\mathbf{R}_h) = -1/3$ are labeled as open circles.

Figure 11. Chiral Worms. Spin variables with $S(\mathbf{R}_t) = +1$ are shown as shaded triangles. Endpoints, branch-points and bends of the domains are decorated by fractional-defect dipoles. Chiral Worms become more prominent compared with Chiral Droplets when the chiral field is large. Fractional defect charge is labeled as in Fig. 10.

Figure 12. Phase diagram for $T_\chi < T_{1/3} < T_1$ and small $f$ (schematic). Along $f = 0$, broken chiral symmetry is restored at the Ising critical point, $T_\chi$. Below $T_\chi$, the Wigner crystal is chiral and has a periodicity $d$. Above $T_\chi$, the Wigner crystal is achiral and has a reduced periodicity of $d/3$. At $T = T_{1/3}$, the achiral Wigner crystal melts by the unbinding of fractionally-charged topological defects. Along the dashed lines, the order parameter correlation function $g_{ij}$ associated with the primary counterion density mode at $G = 2\pi/d$ changes from algebraic decay to exponential decay $g_{ij}$. This change in correlations is associated with the percolation of chiral domain walls that span the system.

Figure 13. External "current" flowing from $i$ to $j$ that enters the computation of the primary correlation function $g_{ij}$ in the dual representation. The hexagonal sites bordering the receive contributions to $\zeta'(\mathbf{R}_h)$ from neighboring gray triangles. An edge-sharing triangle contributes $+2\zeta(\mathbf{R}_t)$, while a non-edge-sharing triangles contributes $-\zeta(\mathbf{R}_t)$ to $\zeta'(\mathbf{R}_h)$ (see text). The resulting distribution of $\zeta'(\mathbf{R}_h)$ used in Eqs. (D.4) and (D.6) is shown in the hexagonal cells.

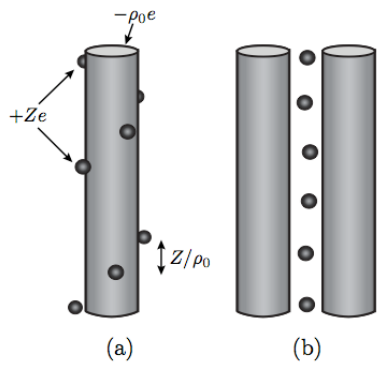

**Figure 1**

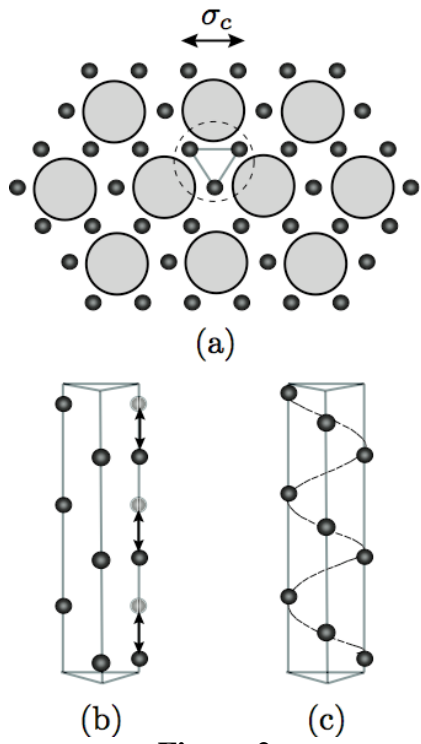

**Figure 2**

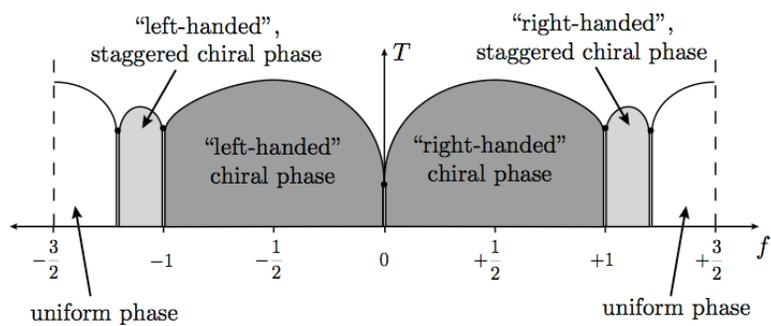

**Figure 3**

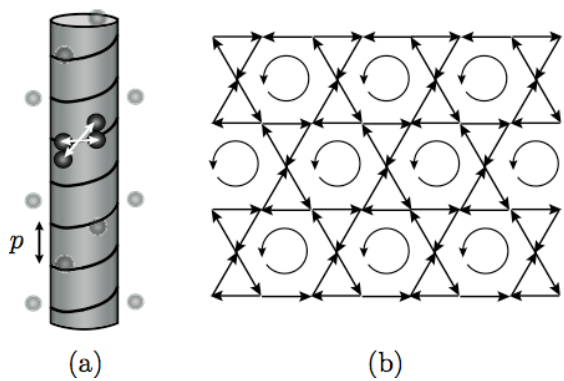

**Figure 4**

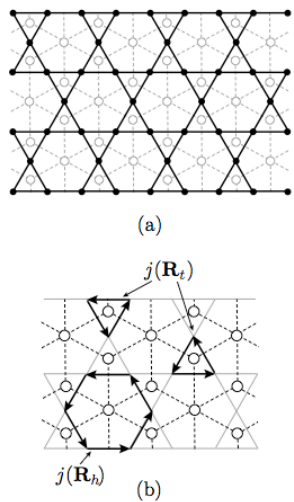

**Figure 5**

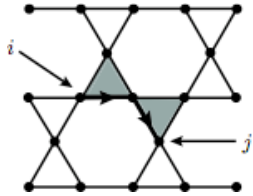

**Figure 6**

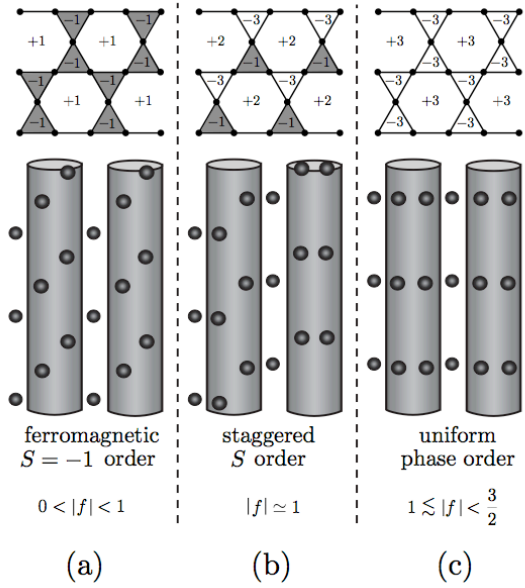

**Figure 7**

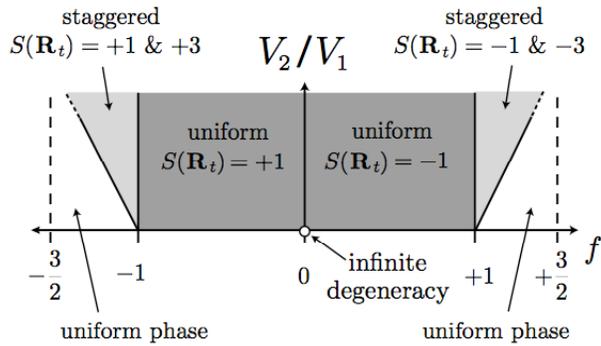

**Figure 8**

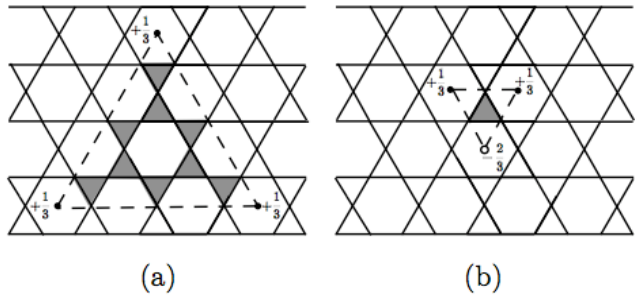

Figure 9

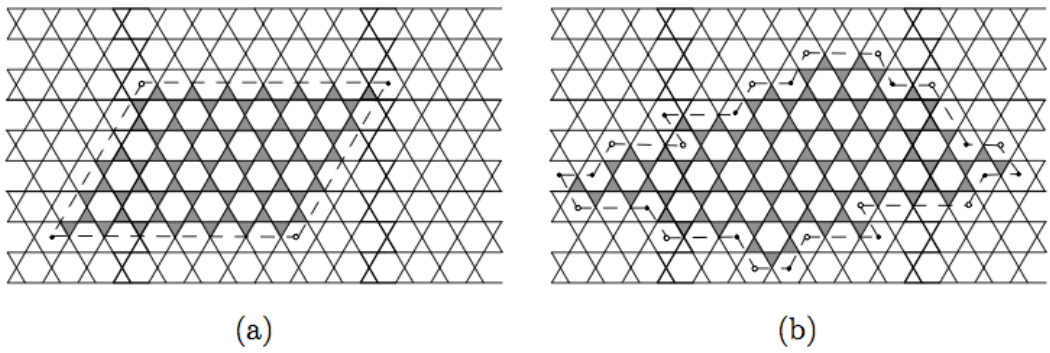

Figure 10

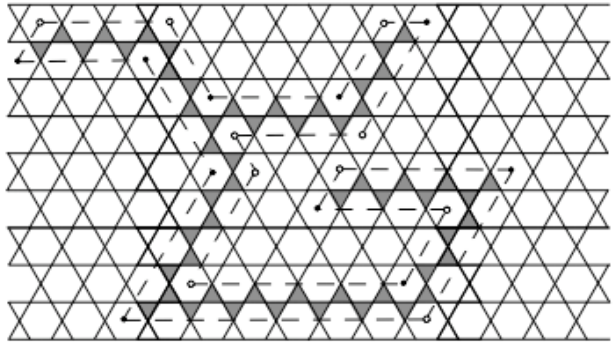

**Figure 11**

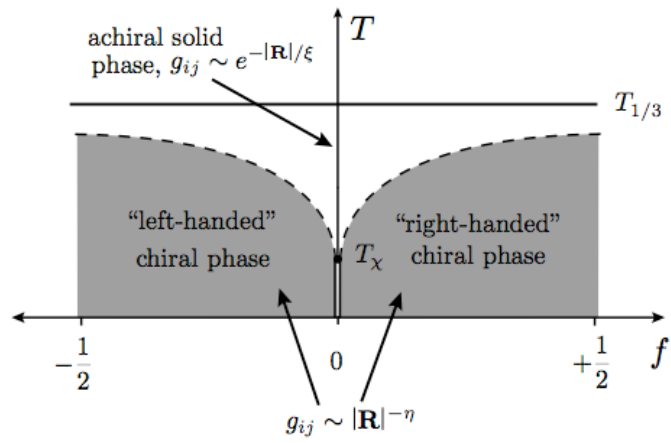

**Figure 12**

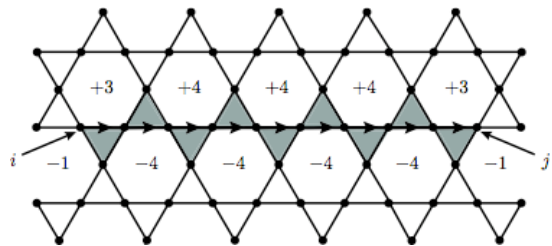

**Figure 13**